\documentclass[12pt]{article}
\usepackage{amsfonts,amssymb}

\def\f{\frac}
\def\s{\sqrt}

\begin{document}

\title{Graviton mass and  total relative density of mass
$\Omega_{\mbox{tot}}$ in  Universe}
\author{S.S.~Gershtein\footnote{gershtein@mx.ihep.su},
A.A.~Logunov\footnote{logunov@mx.ihep.su}, M.A.~Mestvirishvili}
\date{}
\maketitle
\begin{abstract}
It is noticed that the total relative density of mass in the Universe
$\Omega_{\mbox{tot}}$ should exceed $1$, i.e.
$\Omega_{\mbox{tot}}=1+f^2/6$ according to the field relativistic
theory of gravity (RTG), which is free of the cosmological
singularity and which provides the Euclidean character for the
3-dimensional space. Here $f$ is the ratio of the graviton mass $m_g$
to the contemporary value of the ``Hubble mass''
$m^0_H=\hbar H_0/c^2\simeq 3,8\cdot 10^{-66}h$(g)
($h=0,71\pm0,07$). Applying results of the experimental data
processing presented in~[1] an upper limit for the graviton mass is
established as $m_g\leq 3,2\cdot 10^{-66}$g at the 95\% confidence
level.
\end{abstract}

Experimental data obtained in measuring angular characteristics of
the cosmic
microwave background (CMB) spectrum give
 us an opportunity to determine some cosmological parameters~[2],
 including
 $\Omega_{\mbox{tot}}=\rho_{\mbox{tot}}/\rho_c$, which is the ratio
 of
 density of all the matter species $\rho_{\mbox{tot}}$ to the
 critical
 density $\rho_c=3H^2_0/8\pi G$, where $G$ is the gravitational
 constant and
 $H_0$ is the contemporary value of the Hubble constant~[3].
The joint analysis of two recent experiments~BOOMERANG-98~[4] and
MAXIMA-1~[5] is provided in the article by A.~Jaffe et al~[1] with
account
for data of the previous experiment~COBE DMR~[6], and also with
account for
data obtained in studying supernova 1a (SN1a)~[7] and large-scale
structures
of the Universe~(LSS)~[8]. Results of this analysis~[1] show that the
average
value of~$\Omega_{\mbox{tot}}$ systematically exceeds
$\Omega_{\mbox{tot}}=1$
for the combination of different experiments and at the confidence
level 68\%
we have
\begin{equation}
\Omega_{\mbox{tot}} =1,11\pm 0,7\quad (68\%C.L).
\end{equation}
This fact seems very designing. Indeed, according to the most popular
now
theory of inflationary expansion of the Universe at the early
evolution stage
 $\Omega_{\mbox{tot}}$ should be  equal to $1$ with the great
 accuracy (that
 guarantees flatness for the spatial geometry). Therefore, though
$\Omega_{\mbox{tot}}$ equals
\begin{equation}
\Omega_{\mbox{tot}} =1,11^{+0.13}_{-0.12}\quad (95\%C.L)^{[1]}
\end{equation}
in the extended error interval and this does not contradict the
inflationary
expansion model, it is important to note that according to the field
relativistic theory of gravitation (RTG)~[9], whose equations provide
the
Euclidean character of the
3-dimensional space, value of $\Omega_{\mbox{tot}}$ should obligatory
exceed
$1$, and this fact has a fundamental meaning.

 RTG begins with the statement that the source of the gravitational
 field is
 the \underline{conserved} (in Minkowski space) \underline{total}
 energy-momentum tensor, which incorporates the gravitational
 contribution.
 In this respect RTG is similar to modern gauge theories of
 electroweak
 interactions and QCD, where conserved currents are sources for the
 vector
 fields. It follows from the mentioned approach, that the
 gravitational field
 should be described by symmetric tensor of second rank
 $\varphi^{\mu\nu}$.
 This fact, in its turn, leads to the opportunity of a geometrization
 of the
 theory taking into account the universal character of the
 gravitational
 field. Then the motion of matter caused by the gravitational field
 in
 Minkowski space appears as a motion in the effective Riemannian
 space with
 the metric tensor density $\tilde g^{\mu\nu}$, determined by the
 following
 equation
\begin{equation}
\tilde g^{\mu\nu}= \tilde \gamma^{\mu\nu}+
\tilde\varphi^{\mu\nu};\quad
\tilde g^{\mu\nu}= \s{-g}g^{\mu\nu};\quad
g=\det (\tilde g^{\mu\nu})=\det (g_{\mu\nu}),
\end{equation}
where $\tilde \gamma^{\mu\nu}$ and $\tilde \varphi^{\mu\nu}$  are
metric
tensor density and gravitational field tensor density in Minkowski
space.
We are to stress that the mentioned field approach necessarily
requires
nonzero graviton mass  $m_g$, because in the opposite case the metric
tensor
of  Minkowski space disappears from the field equations and we stay
with the
Riemannian space only.
The gravitational field system of equations takes in  RTG the
following
form~[9]
\begin{equation}
\left (
R^\mu_\nu - \f{1}{2} \delta^\mu_\nu R
\right )
+\f{1}{2}
\left (
\f{m_gc}{\hbar}\right )^2
\left (\delta^\mu_\nu+g^{\mu\alpha}\gamma_{\alpha\nu}
-\f{1}{2}
\delta^\mu_\nu g^{\alpha\beta}\gamma_{\alpha\beta}\right )=\f{8\pi
G}{c^2}T^\mu_\nu,
\end{equation}
\begin{equation}
D_\mu \tilde g^{\mu\nu}=0,
\end{equation}
where $R^\mu_\nu$ and $R$ are  corresponding curvatures of the
``effective'' Riemannian space, $T^\mu_\nu$ is the energy-momentum
tensor of
matter in this space, and $D_\mu$ is the covariant derivative in the
Minkowski space. Equations (4)  and
(5) are covariant under arbitrary coordinate transformations and
form-invariant under Lorentz transformations. They correctly describe
all the
gravitational effects, observed in the Solar system.

It is important to mention, that equation (5) arises in this approach
not as
a supplementary condition, but as a consequence of the gravitational
field
equations and the total energy-momentum conservation law.

After taking the interval of the effective Riemannian space for the
homogeneous and isotropic Universe in the following  form
\begin{equation}
ds^2=U(t) dt^2-V(t)
\left [
\f{dr^2}{1-kr^2}+r^2
(d\Theta^2+\sin^2\Theta d\Phi^2)\right ],
\end{equation}
(where $k=1,-1,0$ for the closed, hyperbolic and ``flat'' Universe
respectively) we obtain from equations (5)
\begin{equation}
\f{\partial}{\partial t}\s{\f{V^3}{U}}=0,\;\;\mbox{i.e.}\;\;
V=\beta U^{1/3},\;\; \beta=\mbox{const},
\end{equation}
\begin{equation}
\f{\partial}{\partial r}
[r^2(1-kr^2)^{1/2}]-2r
(1-kr^2)^{-1/2}=0.
\end{equation}
Equation (8) is true only if $k=0$. Therefore, according to  RTG,
\underline{the spatial} \underline{geometry of the Universe should be
flat}
for all the stages of its evolution (irrespective of the fact whether
the
inflationary expansion stage does take place or does not).

After transforming to the proper time $cd\tau=U^{1/2}dt$~~ and
denoting
$a^2(\tau)=U^{1/3}$, we can express interval (6) in the following
form
\begin{equation}
ds^2=c^2d\tau^2-\beta a^2 (\tau)
[dr^2+r^2(d\Theta^2+\sin^2\Theta d\Phi^2)].
\end{equation}
Equations (4) take the following form for scale factor $a(\tau)$
\begin{equation}
\left (
\f{1}{a}
\f{da}{a\tau}
\right )^2
=\f{8\pi G}{3}\rho
-\f{\omega}{a^6}
\left (
1-\f{3a^4}{\beta}+2a^6
\right  ); \quad
\omega =\f{1}{12}
\left (
\f{m_gc^2}{\hbar}
\right )^2;
\end{equation}
\begin{equation}
\f{1}{a}\f{d^2a}{d\tau^2}
=-\f{4\pi G}{3}
\left (
\rho+ \f{3P}{c^2}
\right )- 2\omega
\left (
1-\f{1}{a^6}
\right ).
\end{equation}
The constant $\beta$, appearing in expressions (9)-(10), has a simple
meaning. According to the causality principle, the matter motion in
the
Minkowski space could not be processed outside the causality cone of
this
space. This leads to the following condition
$a^4 (\tau) \leq \beta$, i.e. $\beta=a^4_{\mbox{max}}$.

Thus, according to  RTG, the expansion of the Universe can not be
unlimited.
So, this excludes a possibility to explain the accelerated expansion
of the
Universe, which is now observed, on the base of the assumption of a
nonzero
vacuum energy.
This looks quite natural because the vacuum energy density should be
identically zero in the field treatment of the gravitational field.
From the viewpoint of RTG an assumption of existence of a special
substance
called quintessence~[10] is quite suitable for an explanation of the
observed
acceleration. Its equation of state should be of the following form
\begin{equation}
P=-(1-\nu)\rho_q;\;\;
(0<\nu<2/3),
\end{equation}
and its density should decrease according to $\rho_q\sim
1/a^{3\nu}$. (This opportunity has been pointed  out by
Kalashnikov~[11].) It
follows from equation (10), that the maximal expansion of the
Universe $\left
( \f{da}{d\tau}=0\;\;\mbox{¨}\;\;
a>>1\right)$ is reached at density
$\rho_{\min}=\f{1}{16\pi G}
\left (
\f{m_gc^2}{\hbar}\right )^2$.
From equation (10) it is also evident, that according to RTG the
cosmological
singularity is also excluded, because, due to the positive
definiteness of
l.h.s. of the equation, the increase of the matter density at the
radiation-dominant stage as $\rho\sim 1/a^4$ under $a\to 0$ should
compensate
the negative term in  r.h.s., increasing as  $(\omega/a^6)$, at some
minimal
value  $a_{\mbox{min}}\neq =0$.

Thus, the nonzero graviton mass in  RTG leads to the cyclewise
character of
the Universe evolution, excluding both
the cosmological singularity, and the possibility of an unlimited
expansion
of the Universe, and providing also
\underline{at all the stages of evolution} the flat geometry for
3-dimensional space. However, we have
$\Omega_{\mbox{tot}}> 1$. Indeed, by writing equation (10) for the
contemporary stage $(a\gg 1)$ and  dividing
both sides of it by the Hubble constant $H^2_0$, we obtain 
\begin{equation}
\Omega_{\mbox{tot}} =1+f^2/6; \quad f=\f{m_g
c^2}{\hbar  H_0}=\f{m_g}{m^0_H} \end{equation}where $m^0_H$ could be called
``Hubble
mass'': 
\begin{equation}
m^0_H=\f{\hbar H_0}{c^2}=h
3.8\cdot 10^{-66}\mbox{g}\simeq 2.7\cdot 10^{-66}\mbox{g}\;\;
(h=0.71), 
\end{equation}
From (2) and (13) it follows
$f^2/6\leq 0.24$, i.e. 
\begin{equation}
m_g\leq 1.2 m^0_H=3.2\cdot
10^{-66}\mbox{g}\;\;
(95\%C.L). 
\end{equation}
If we use  (1), then
$0.04\leq \f{f^2}{6}\leq 0.18$ and, consequently, the graviton mass
can be
estimated as follows 
\begin{equation}
m_g=2.2^{+0.6}_{0.9}\cdot 10^{-66}\mbox{g}\;\; (68\%C.L) 
\end{equation}
The closeness of
values $m_g\simeq m^0_H$ rises a
question: is this coincidence accidental or it has a more fundamental
nature?
It should be noticed that the
estimates obtained here do not contain any model suggestions (besides
the
hypothesis that the gravitational field
could be treated as a physical field in Minkowski space and its
source is the
total energy-momentum tensor,
incorporating the gravitational field contribution).

The authors express their sincere gratitude to V.A.~Petrov,
N.P.~Tkachenko
and N.E.~Tyurin for fruitful discussions. One of the authors (S.S.G.)
is
grateful for support to RFBR (Grants No. 01-02 16585 and 00-15
96645).

{\bf Note added.} Since  this article had been sent to the journal
Doklady
Akademii Nauk, preliminary results from WMAP appeared
(arXiv:astro-ph/0302207
of February 12, 2003) which presented data of  more precise
measurements:
$\Omega_{tot}=1.02\pm 0.02$. These data permit to strengthen the
upper limit
for the graviton mass. At the level of $2\sigma$ $f^2/6\le 0.06$,
i.e.
$m_g\le 0.6m_H^0=1.6\cdot 10^{-66}$ g.

\enddocument